\documentclass[%
 aip,
 amsmath,amssymb,
preprint,%
]{revtex4-1}

\usepackage{graphicx}% Include figure files
\usepackage{dcolumn}% Align table columns on decimal point
\usepackage{bm}% bold math
\usepackage[utf8]{inputenc}
\usepackage[T1]{fontenc}
\usepackage{mathptmx}
\usepackage{etoolbox}
\usepackage{yfonts}
\usepackage{xcolor}

\usepackage{soul}  % Strikeout text: \st{bad text}

\usepackage[colorlinks=true, allcolors=blue]{hyperref}
\usepackage{mathtools}

\newcommand\ddfrac[2]{\frac{\displaystyle #1}{\displaystyle #2}}
\newcommand{\Xrm}{\mathrm{X}}
\newcommand{\ra}{r_\mathrm{a}}
\newcommand{\rd}{r_\mathrm{d}}
\newcommand{\rdiff}{r_\mathrm{diff}}

\makeatletter
\def\@email#1#2{%
 \endgroup
 \patchcmd{\titleblock@produce}
  {\frontmatter@RRAPformat}
  {\frontmatter@RRAPformat{\produce@RRAP{*#1\href{mailto:#2}{#2}}}\frontmatter@RRAPformat}
  {}{}
}%
\makeatother

\begin{document}

\preprint{AIP/123-QED}

\title{Surface Coverage Dynamics for Reversible Dissociative Adsorption \\ on Finite Linear Lattices}

\author{Enrique Mercado}
\affiliation{Department of Applied Mathematics, University of California, Merced, California 95343, USA}

\author{Hyun Tae Jung}
\affiliation{Department of Chemistry, Korea Advanced Institute of Science and Technology, Daejeon 34141, South Korea}

\author{Changho Kim}
\email{ckim103@ucmerced.edu}
\affiliation{Department of Applied Mathematics, University of California, Merced, California 95343, USA}

\author{Alejandro L. Garcia}
\affiliation{Department of Physics and Astronomy, San Jose State University, San Jose, California 95192, USA}

\author{Andy J. Nonaka}
\affiliation{Center for Computational Sciences and Engineering, Lawrence Berkeley National Laboratory, Berkeley, California 94720, USA}

\author{John B. Bell}
\affiliation{Center for Computational Sciences and Engineering, Lawrence Berkeley National Laboratory, Berkeley, California 94720, USA}

\date{\today}% It is always \today, today,

\begin{abstract}
Dissociative adsorption onto a surface introduces dynamic correlations between neighboring sites not found in non-dissociative absorption.
We study surface coverage dynamics where reversible dissociative adsorption of dimers occurs on a finite linear lattice.
We derive analytic expressions for the equilibrium surface coverage as a function of the number of reactive sites, $N$, and the ratio of the adsorption and desorption rates.
Using these results, we characterize the finite size effect on the equilibrium surface coverage.
For comparable $N$’s, the finite size effect is significantly larger when $N$ is even than when $N$ is odd.
Moreover, as $N$ increases, the size effect decays more slowly in the even case than in the odd case.  
The finite-size effect becomes significant when adsorption and desorption rates are considerably different.
These finite-size effects are related to the number of accessible configurations in a finite system where the odd-even dependence arises from the limited number of accessible configurations in the even case.
We confirm our analytical results with kinetic Monte Carlo simulations.
We also analyze the surface-diffusion case where adsorbed atoms can hop into neighboring sites. 
As expected, the odd-even dependence disappears because more configurations are accessible in the even case due to surface diffusion.
\end{abstract}

\maketitle

\section{\label{sec_intro}Introduction}

Adsorption and desorption processes provide an essential mechanism for mass transport at fluid-solid interfaces~\cite{Masel2003, OuraLifshitsSaraninZotovKatayama2003}.
For example, heterogeneous catalysis~\cite{Schloegl2015, MateraReuter2010} relies on this mechanism to transport reactants from a gas onto a catalytic surface where chemical reactions occur and bring products back to the gas phase.
Therefore, developing a correct description of adsorption and desorption processes is a crucial step in the computational modeling of gas-solid interfacial systems requiring appropriate modeling assumptions and careful analysis~\cite{StamatakisVlachos2012, LiuEvans2013, NagasakaKondohNakaiOhta2007, PiccininStamatakis2014}.
While the inclusion of lateral interactions (also referred to as adsorbate-adsorbate interactions) is important for realistic modeling of the phenomenon, it makes the analytic investigation of the behavior of the resulting system intractable.
In this paper, we consider a theoretical model of reversible dissociative adsorption based on Langmuir adsorption modeling and show that, even without lateral interactions, the phenomenon of reversible dissociative adsorption exhibits rich dynamics that requires detailed analysis.

The Langmuir adsorption model~\cite{Masel2003, OuraLifshitsSaraninZotovKatayama2003, SwensonStadie2019} has served as the most influential theoretical model for reversible adsorption processes.
Despite its simplicity, the model not only captures key molecular features but also gives analytical expressions for adsorption isotherms.
One of the fundamental assumptions of the model is that there are no interactions between adsorbates on adjacent sites.
This assumption implies another assumption, namely, that the occupancy of each site becomes uncorrelated in the \textit{infinite equilibrium system}, from which one can derive the Langmuir isotherms for both non-dissociative (or one-site) and dissociative (or two-site) adsorption.
However, we note that, for the dissociative adsorption case, the validity of the uncorrelated site occupancy assumption may not be guaranteed in a general situation (i.e.\ finite or nonequilibrium system) because adsorption and desorption events can lead to dynamic correlations between neighboring sites.
These kinetically induced correlations present in the dissociative adsorption case make the analysis of system behaviors nontrivial.
This is contrasted with the non-dissociative adsorption case, where the uncorrelated site occupancy assumption always holds and the system exhibits trivial exponential kinetics because the occupancy of each site can be modeled by an independent stochastic process under the assumption of no lateral interactions.

Theoretical investigations of kinetically induced correlations in a system undergoing two-site adsorption events date back to at least the 1939 study by Flory~\cite{Flory1939} where intramolecular reactions on polymer chains were modeled by irreversible two-site adsorption events (i.e.\ without desorption events).
This study has led to various analyses of emerging behavior in the random and cooperative sequential adsorption models~\cite{Evans1993}.
For the reversible dissociative adsorption case, a systematic analysis was performed in the context of deposition and evaporation of $k$-mers ($k=2,3,\cdots$) on a linear lattice~\cite{BarmaGrybergStinchcombe1993, StinchcombeGrynbergBarma1993}.
Using an equivalent formulation based on a quantum-spin model, it was shown that the autocorrelation function for the number of adsorbed atoms exhibits a power-law decay in time (i.e.\ $\sim t^{-1/2}$).
To this end, in the context of the Goldstone theorem~\cite{GoldstoneSalamWeinberg1962}, a family of conservation laws was identified by dividing the lattice into $k$ sublattices and considering the number of atoms adsorbed on each sublattice.
While this approach provides an insightful explanation for the origin of the nontrivial power-law decay observed in the infinite-system limit, we note that the approach is not fully applicable to a finite system, particularly if the size of the lattice is not a multiple of $k$.
Liu and Evans~\cite{LiuEvans2015} analyzed spatial correlations in one- or two-dimensional lattice systems undergoing reversible dissociative adsorption using a similar formulation with sublattices.
They showed that the magnitude of nearest-neighbor and other short-range correlations decay like $t^{-d/2}$ where $d$ is the dimensionality of the lattice.
This analysis for the dimer ($\mathrm{O_2}$) adsorption-desorption process was performed as part of a broader study of a $\mathrm{CO}$ oxidation model~\cite{ReuterScheffler2006} and strong spatial correlations appearing in some quasi-steady states during the evolution of surface coverage dynamics were used to explain why phenomenological kinetics (e.g.\ mean-field description) may fail to provide an adequate description of heterogeneous catalysis~\cite{TemelMeskineReuterSchefflerMetiu2007, MateraMeskineReuter2011}.
We note that previous studies~\cite{BarmaGrybergStinchcombe1993, StinchcombeGrynbergBarma1993, LiuEvans2015} mostly considered the infinite-system limit and analyzed the power-law decay of dynamic correlations to demonstrate the intriguing nature of the reversible dissociative adsorption dynamics.

In this paper, we perform a systematic analysis of reversible dissociative adsorption occurring on a \textit{finite} linear lattice.
The main quantity of interest is the equilibrium surface coverage.
We show that, contrary to the infinite system case, the effect of kinetically induced correlations can be seen in this \textit{static} quantity because the occupancy of each site is not completely uncorrelated in a finite system. 
We also show that the finite system-size effect on the equilibrium surface coverage exhibits interesting behavior depending on whether the lattice has an odd or even number of sites and this can be explained by the number of accessible configurations that a finite system can have.
Analyzing finite system-size effects and investigating their physical origins are crucial to understanding emerging behaviors in finite-sized physical systems (e.g.\ nano-engineered materials)~\cite{YuanDoanGrabowBrankovic2017, EhrlingMiuraSenkovskaKaskel2021} or perform reliable simulation studies using finite-sized computational models (e.g.\ with periodic boundary conditions)~\cite{Frenkel2013, KimHanKimLiKarniadakisLee2018, KimKimKarniadakisLeeKozak2019, LeeKimColvin2022}.
In fact, considering that even high-quality single crystal surfaces have terraces that seldom exceed a size of a few hundred sites in one direction, we note that most real catalytic systems are nano-structured without any particular engineering.

We formulate a continuous-time Markov chain model~\cite{Anderson1991} for a linear lattice undergoing reversible dissociative adsorption and investigate the dynamics of this system using the following two methods.
First, we develop an analytic approach based on the chemical master equation (CME).
The CME is a set of first-order differential equations that describe the probabilistic time evolution of the system in state space~\cite{Makarov2017, KulasiriKosarwal2021}.
We derive analytical expressions for the equilibrium surface coverage as a function of system size and the ratio of the adsorption and desorption rates.
Second, we perform lattice kinetic Monte Carlo (KMC) simulations~\cite{Jansen2012, AndersenPanosettiReuter2019, PinedaStamatakis2022} to confirm the validity of our analytical results for the equilibrium surface coverage.
KMC numerically solves the CME in the sense that the distribution of sample trajectories is the solution of the CME. 
Furthermore, we investigate the time evolution of the surface coverage dynamics to discuss how the equilibrium surface coverage is reached.
In addition, using both CME and KMC approaches, we investigate the effect of surface diffusion, which is known to reduce spatial correlations~\cite{LiuEvans2015, StamatakisVlachos2011}.

The rest of the paper is organized as follows.
In Section~\ref{sec_sys}, we introduce our lattice system and reversible dissociative adsorption and derive the Langmuir isotherm.  
In Sections~\ref{sec_cme}, we formulate a continuous-time Markov chain model and derive analytic results for the equilibrium surface coverage and the correlation coefficient for the occupancy of neighboring sites.
Using these results as well as KMC simulations, we analyze finite system-size effects on equilibrium surface coverage.
In Section~\ref{sec_surfdiff}, we analyze the effect of surface diffusion.
In Section~\ref{sec_conclusion}, we conclude the paper with a summary and an outline for future work.

%%%%%%%%%%%%%%%%%%%%%%%%%%%%%%%%%%%%%%%%%%%%%%%%%%%%%%%%%%%%%%%
\section{\label{sec_sys}System}

\begin{figure}
\includegraphics[width=0.8\linewidth]{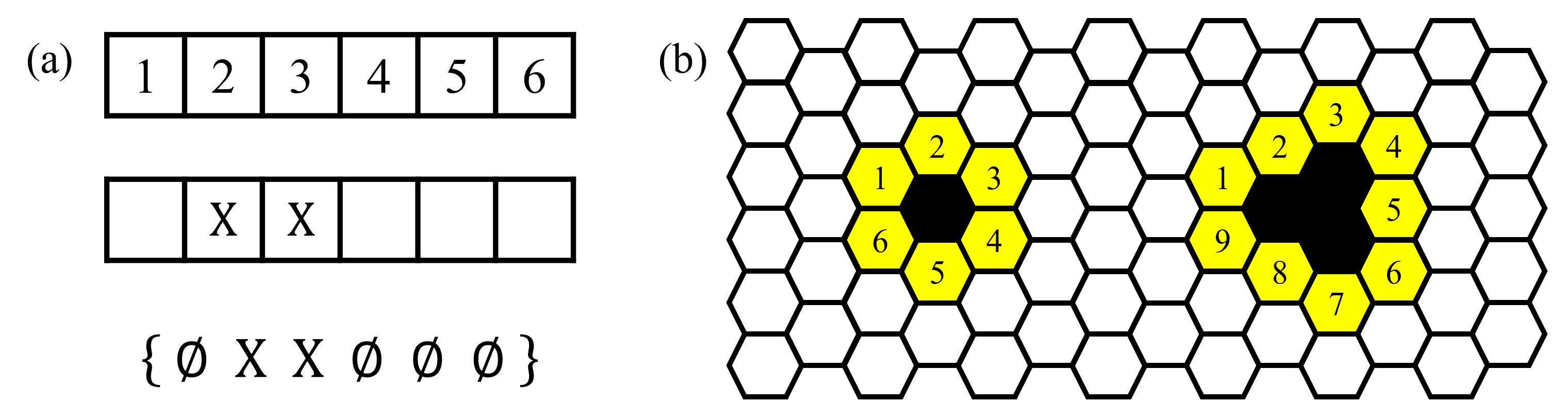}
    \caption{\label{fig_lattice}
    (a) An example of a linear lattice with $N=6$ sites and a typical configuration with 2 occupied sites (marked with an X).
    (b) A linear lattice with a periodic boundary can be viewed as a simple theoretical model for active sites (in yellow) adjacent to doped sites (in black) on a surface of inert sites (in white). Two cases with $N=6$ and 9 are shown.}
\end{figure}

We consider a theoretical model of reversible dissociative adsorption based on Langmuir adsorption modeling.
The surface is represented as a linear lattice with $N$ reactive sites, see Figure~\ref{fig_lattice}(a).
We assume that all reactive sites are identical and each site has two neighboring sites.
In other words, the terminal sites are connected via a periodic boundary.
One can consider this lattice system with a sufficiently large number of sites as an approximation to an infinite lattice.
One may also view this linear lattice model with a small number of sites as a simple theoretical model for active sites adjacent to doped sites~\cite{HakalaKronbergLaasonen2017}, see Figure~\ref{fig_lattice}(b).

As a reaction model, we consider reversible dissociative adsorption of diatomic gas molecules $\Xrm_2$ between the gas phase and the lattice:
\begin{equation}
\label{reaction}
    \varnothing\varnothing + \Xrm_2 (\mathrm{g}) \xrightleftharpoons[\rd]{\ra} \Xrm\Xrm.
\end{equation}
A molecule $\Xrm_2$ in the gas phase dissociates into atoms that are adsorbed onto two neighboring empty sites (denoted as $\varnothing\varnothing$) and vice versa.
The rates of the forward reaction (dissociative adsorption) and the reverse reaction (associative desorption) are denoted as $\ra$ and $\rd$, respectively.
$\ra$ and $\rd$ have units of inverse time and are taken as constant.
Note that the dependence on the partial pressure of $\Xrm_2$ is included in $\ra$ and no lateral interactions are assumed.
In addition we will consider surface diffusion of the adsorbed atom $\Xrm$ in Section~\ref{sec_surfdiff}, see Eq.~\eqref{surfdiff}.
The main quantity of interest in this paper is the surface coverage of the system.
The surface coverage $\theta_N(t)$ is defined as the ratio of the mean number of occupied sites at time $t$ to the number of reactive sites, $N$.
We assume that the lattice is initially unoccupied, i.e.\ $\theta_N(0)=0$.
We define the equilibrium surface coverage $\bar{\theta}_N$ as the long-time (or steady-state) limit of $\theta_N(t)$, i.e.\ $\bar{\theta}_N = \lim_{t\rightarrow\infty}\theta_N(t)$.

Before closing this section, we derive the Langmuir isotherm, which is the equilibrium surface coverage in the infinite system, i.e.\ $\bar{\theta}_\infty = \lim_{N\rightarrow\infty}\bar{\theta}_N$.
We let brackets denote the probability that a certain $n$-site cluster ($n=1,2,3$) is found in the lattice system.
Specifically, $[\Xrm]$ and $[\varnothing]$ denote the probabilities that a site is occupied by $\Xrm$ and unoccupied, respectively, whereas $[\Xrm\Xrm]$, $[\Xrm\varnothing]$, $[\varnothing\Xrm]$, and $[\varnothing\varnothing]$ denote the probabilities that a nearest-neighbor pair of sites is in states of 
$\Xrm\Xrm$, $\Xrm\varnothing$, $\varnothing\Xrm$, and $\varnothing\varnothing$, respectively.
By the hierarchical rate equations, the time evolution of $[\Xrm]$ is given as~\cite{LiuEvans2015}
\begin{equation}
\label{dXdt}
    \frac{d}{d t}[\Xrm] = 2 \ra [\varnothing\varnothing] - 2 \rd [\Xrm\Xrm].
\end{equation}
Although Eq.~\eqref{dXdt} is exact, it cannot be solved as an initial value problem because the equation is not closed.
The two-site cluster quantities, $[\varnothing\varnothing]$ and $[\Xrm\Xrm]$, are not determined by the one-site cluster quantities, $[\Xrm]$ and $[\varnothing]=1-[\Xrm]$.
However, assuming an infinite equilibrium system, one can derive the steady-state value of $[\Xrm]$, which is $\bar{\theta}_\infty$, from Eq.~\eqref{dXdt}.
Since we assume there are no lateral interactions, the occupancy of each site becomes uncorrelated in the thermodynamic equilibrium~\cite{SwensonStadie2019} and thus one has $[\Xrm\Xrm]=[\Xrm][\Xrm]=\bar{\theta}_\infty^2$ and $[\varnothing\varnothing]=[\varnothing][\varnothing] = (1-\bar{\theta}_\infty)^2$.
By combining these with the condition that $\frac{d}{dt}[\Xrm]$ becomes zero in the steady state, one finally obtains~\cite{SwensonStadie2019}
\begin{equation}
\label{thetaeq_inf_sec2}
    \bar{\theta}_\infty = \frac{1}{1+\sqrt{k}},\quad\mbox{where}\quad k = \frac{\rd}{\ra}.
\end{equation}
Note that Eq.~\eqref{thetaeq_inf_sec2} cannot be used to derive any finite-system or nonequilibrium results, such as $\theta_N(t)$, $\bar{\theta}_N$, or $\theta_\infty(t)$, because the uncorrelated site occupancy assumption is not guaranteed to hold and $[\mathrm{AB}]\neq[\mathrm{A}][\mathrm{B}]$ in general.

%%%%%%%%%%%%%%%%%%%%%%%%%%%%%%%%%%%%%%%%%%%%%%%%%%%%%%%%%%%%%%%
\section{\label{sec_cme}Finite-Size Effects: Even \& Odd $N$}

In this section, we develop a continuous-time Markov chain description for the system described in Section~\ref{sec_sys} and investigate finite system-size effects on equilibrium surface coverage $\bar{\theta}_N$ using both analytical and simulation approaches.
The analytical approach is based on the master equation description, for which a complete characterization of all accessible configurations is a prerequisite.
As briefly mentioned in the Introduction, the finite-size effect is related to the number of accessible configurations.
We present a combinatorial argument to characterize all accessible configurations in Section~\ref{subsec_conf}.
We then describe our master-equation-based approach in Section~\ref{subsec_cme_des}.
We present analytic results for the equilibrium surface coverage and the correlation coefficient for the occupancy of neighboring sites in Sections~\ref{subsec_surfcov} and \ref{subsec_corr}, respectively.
We finally present KMC simulation results in Section~\ref{subsec_kmc}.

%%%%%%%%%%%%%%%%%%%%%%
\subsection{\label{subsec_conf}Accessible Configurations}

As described in Section~\ref{sec_sys}, we consider a periodic linear strip with $N$ sites and label these sites from 1 to $N$.
We assume that each site in the system is initially unoccupied.
We define a \textit{configuration} of the system by specifying whether each site is occupied (denoted by $\Xrm$) or unoccupied (denoted by $\varnothing$).
We say that a configuration is \textit{accessible} from another configuration if the former can be obtained from the latter via a sequence of reactions.
For example, for $N=5$, configuration $\{\Xrm\varnothing\varnothing\Xrm\varnothing\}$ is accessible from the initial configuration $\{\varnothing\varnothing\varnothing\varnothing\varnothing\}$ because the former is obtained from the latter via two adsorption events at sites 1-2 and 3-4 followed by a desorption event at sites 2-3.
Using a combinatorial argument, we count the total number of accessible configurations, denoted by $n_\mathrm{tot}$.
We also determine the number of accessible configurations with $2l$ occupied sites, denoted by $\tilde{n}_l$, which will be used to derived $\bar{\theta}_N$ in Section~\ref{subsec_surfcov}.

Since each site is either occupied or unoccupied, there are a total of $2^N$ configurations.
However, it is important to note that all configurations are not accessible to each other.
This is because of properties that must continue to hold when the system undergoes a sequence of dissociative adsorption and associative adsorption events.
We first note that the parity (i.e.\ whether odd or even) of the number of occupied sites does not change because the number of occupied sites increases (or decreases) by 2 via a dissociative adsorption (or associated desorption) event.

For the odd case of $N=2m+1$, this parity can completely characterize the two groups of configurations within which all configurations are accessible from each other.
The $2^{N-1}$ configurations with an even number of occupied sites are accessible from the completely unoccupied configuration $\{\varnothing\varnothing\cdots\varnothing\}$, whereas the other $2^{N-1}$ configurations with an odd number of occupied sites are accessible from the completely occupied configuration $\{\Xrm\Xrm\cdots\Xrm\}$.
Hence, we will consider only the former group of configurations in the continuous-time Markov chain description for the odd case.
One can easily see that the number of configurations with $2l$ occupied sites ($l=0,1,\cdots,m$) is given by $\binom{N}{2l}$ where $\binom{m}{j}=\frac{m!}{j!(m-j)!}$ denotes a binomial coefficient.
The total number of accessible configurations, $n_\mathrm{tot} = 2^{N-1}$, is then given by the sum of $\tilde{n}_l$: 
\begin{equation}
\label{ntotnlodd}
    \tilde{n}_l = \binom{2m+1}{2l}, \quad 
    n_\mathrm{tot} = \sum_{l=0}^{m} \tilde{n}_l = 2^{2m} = 2^{N-1}.
\end{equation}

For the even case of $N=2m$, there is an additional conserved quantity~\cite{BarmaGrybergStinchcombe1993, StinchcombeGrynbergBarma1993, LiuEvans2015}.
To define this quantity, we consider the two alternating sublattices.
The (+) sublattice only has sites with an odd site number, whereas the (-) sublattice only contains sites with an even site number.
We denote the number of occupied sites in the (+) and (-) sublattices by $N_+$ and $N_-$, respectively.
Since dissociative adsorption and associative desorption events occur at two neighboring sites, one of the sites belongs to the (+) sublattice and the other belongs to the (-) sublattice.
As a result, the quantity $N_+ - N_-$ is conserved when adsorption or desorption occurs.
Hence, contrary to the odd case, all $2^{N-1}$ configurations with an even number of occupied sites are not accessible from the initial unoccupied configuration.
For example, for $N=6$, $\{\Xrm\Xrm\Xrm\varnothing\Xrm\varnothing\}$ is not accessible from $\{\varnothing\varnothing\varnothing\varnothing\varnothing\varnothing\}$ because the $N_+ - N_-$ values of the former and latter configurations (2 and 0, respectively) are different.

Combinatorial expressions for $\tilde{n}_l$ and $n_\mathrm{tot}$ in the even case are obtained as follows.
Since we only consider configurations accessible from the initial unoccupied configuration, those configurations have $N_+ - N_- = 0$.
Thus, if a configuration has $2l$ occupied sites (i.e.\ $N_+ + N_- = 2l$), we know $N_+ = N_- = l$.
We note that there are $\binom{m}{l}$ ways to arrange $N_+ = l$ atoms in the (+) sublattice with $m$ sites and the same expression holds for the (-) sublattice.
Hence, we obtain
\begin{equation}
\label{ntotnleven}
    \tilde{n}_l = \binom{m}{l}^2, \quad 
    n_\mathrm{tot} = \sum_{l=0}^{m} \tilde{n}_l = \binom{2m}{m} = \binom{N}{N/2}.
\end{equation}
Note that the conservative quantity, $N_+ - N_-$, was identified using sublattices in previous studies~\cite{BarmaGrybergStinchcombe1993, StinchcombeGrynbergBarma1993, LiuEvans2015}.
However, the main focus in these works was to explain the power-law decay of time-correlation functions observed in the infinite system limit and combinatorial arguments were not developed.

%%%%%%%%%%%%%%%%%%%%%%
\subsection{\label{subsec_cme_des}Continuous-Time Markov Chain Description}

We construct a continuous-time Markov chain model for the system described in Section~\ref{sec_sys} by considering all accessible configurations and defining transition rates between each pair of configurations.
If two configurations are obtained from each other by an adsorption or desorption event, the transition rates are set to $\ra$ and $\rd$, respectively; otherwise, zero transition rates are set.
To describe the time evolution of the continuous-time Markov chain model, one can use the chemical master equation (CME), which is a set of first-order differential equations whose solution gives the probability that the system is in a certain configuration at a certain time.
However, the dimension of the CME, which is equal to $n_\mathrm{tot}$, grows quickly, see Eqs.~\eqref{ntotnlodd} and \eqref{ntotnleven}.
Moreover, even for small values of $N$, the values of $n_\mathrm{tot}$ are rather large (e.g., $n_\mathrm{tot}=16$ and 20 for $N=5$ and 6, respectively), which makes it difficult to investigate the CME analytically.
Instead of the standard approach which keeps track of all accessible configurations separately, we group configurations with the same characteristics into an aggregated state and write the CME for those aggregated states.
Owing to the periodic boundary or the ring structure, configurations obtained via cyclic translation belong to the same aggregated state.
We introduce a notation $\mathbb{C}(\alpha)$ to denote an aggregated state containing the configuration $\alpha$ and all the other configurations that are reached from $\alpha$ via cyclic translation.

\begin{figure}
    \includegraphics[width=0.75\linewidth]{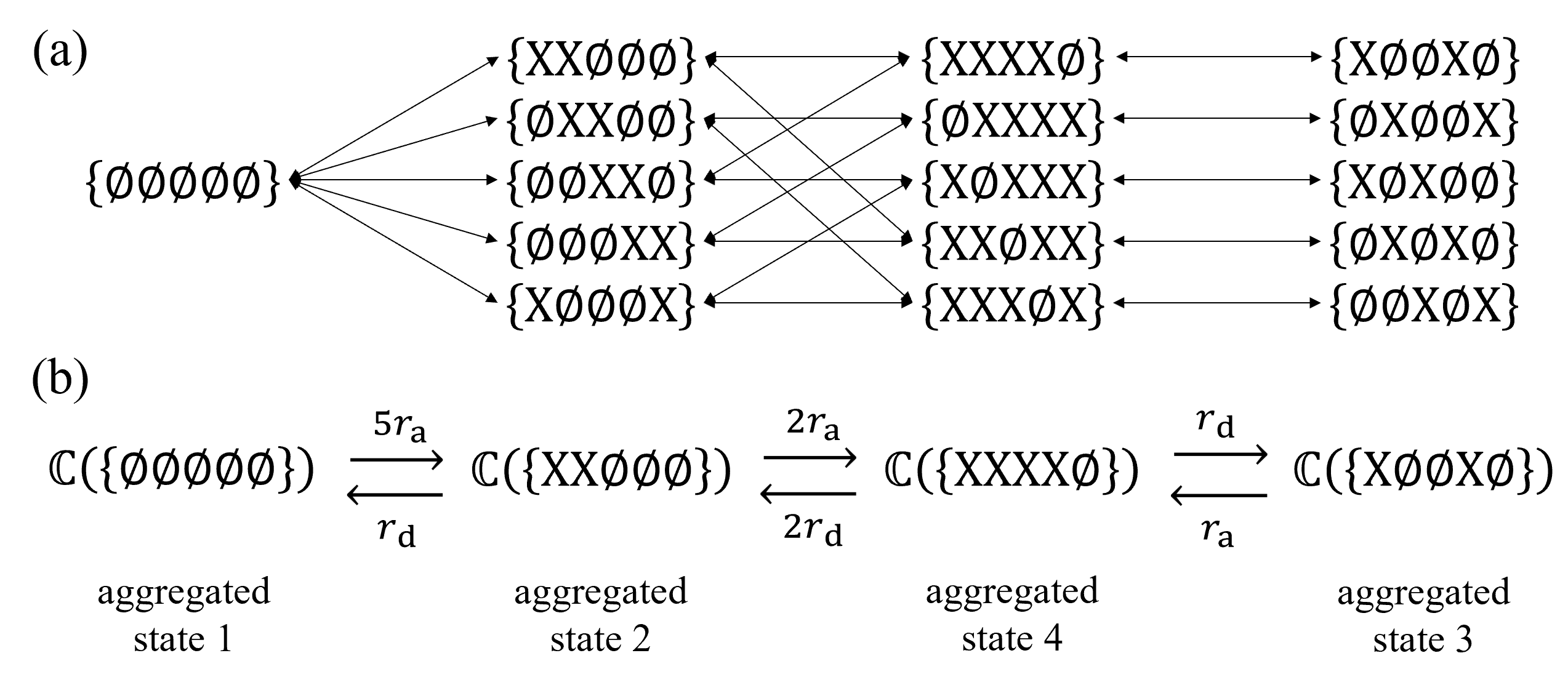}
    \caption{\label{fig_5site}
    Transition diagrams for a 5-site system.
    Panel~(a) shows the standard approach, where 16 configurations accessible from the initial unoccupied configuration are considered separately.
    A double-sided arrow between two configurations indicates that one configuration is obtained from the other configuration via either adsorption or desorption and vice versa.
    The transition rate is $\ra$ (or $\rd$) for the direction increasing (or decreasing) the number of occupied sites.
    Panel~(b) shows our approach based on aggregated states.
    Note that the transition rates between aggregated states are multiples of $\ra$ and $\rd$.
    See the main text for how the multiplicities are determined.}
\end{figure}

We use a 5-site system as an example to explain how to construct the transition diagram and write the CME for aggregated states.
Figure~\ref{fig_5site}(a) shows all 16 accessible configurations and the transition diagram for the standard approach. 
As shown in Figure~\ref{fig_5site}(b), they are grouped into the following 4 aggregated states:
$\mathbb{C}(\{\varnothing\varnothing\varnothing\varnothing\varnothing\})$,
$\mathbb{C}(\{\Xrm\Xrm\varnothing\varnothing\varnothing\})$,
$\mathbb{C}(\{\Xrm\varnothing\varnothing\Xrm\varnothing\})$, and
$\mathbb{C}(\{\Xrm\Xrm\Xrm\Xrm\varnothing\})$.
For each state $i$, we introduce $l_i$ so that each configuration in the state has $2 l_i$ $\Xrm$ atoms.
For the current example, we have $(l_1,l_2,l_3,l_4)=(0,1,1,2)$.
The transition rate from state $i$ to state $i'$ is given as a multiple of $\ra$ or $\rd$ when adsorption (for $l_{i'}=l_i + 1$) or desorption (for $l_{i'}=l_i - 1$) occurs, respectively.
The multiplicity factor is determined by counting how many configurations in state $i'$ can be obtained from each configuration in state $i$.
Thus, the transition rates for $1\rightarrow2$, $2\rightarrow4$, and $3\rightarrow4$ are $5\ra$, $2\ra$, and $\ra$, respectively, whereas the transition rates for $2\rightarrow1$, $4\rightarrow2$, and $4\rightarrow3$ are $\rd$, $2\rd$, and $\rd$, respectively.
Using these transition rates, we obtain the following CME:
\begin{equation}
\begin{split}
    \dot{p}_1 &= -5\ra p_1 + \rd p_2, \\
    \dot{p}_2 &= 5\ra p_1 - (2\ra+\rd) p_2 + 2\rd p_4, \\
    \dot{p}_3 &= -\ra p_3 + \rd p_4, \\
    \dot{p}_4 &= 2\ra p_2 + \ra p_3 - 3\rd p_4,
\end{split}
\end{equation}
where $p_i$ denotes the probability that the system is in state $i$.
Equivalently, we have a matrix form:
\begin{equation}
\label{cme_mform}
    \dot{\mathbf{p}} = R \mathbf{p}.
\end{equation}

Once we have written down a CME~\eqref{cme_mform} for an $N$-site system, its solution can be expressed using a matrix exponential: $\mathbf{p}(t) = e^{tR}\mathbf{p}(0)$. 
While obtaining an analytic expression of $\mathbf{p}(t)$ is not a trivial task even for small $N$, one can determine $\mathbf{p}(t)$ accurately using a numerical method (e.g.\ Runge--Kutta).
The equilibrium probability distribution $\mathbf{p^*}$ is then given as the long-time limit of $\mathbf{p}(t)$, i.e.\ $\mathbf{p^*} = \lim_{t\rightarrow\infty} \mathbf{p}(t)$.
Alternatively, the equilibrium probability distribution $\mathbf{p^*}$ can be obtained as the unique invariant probability distribution satisfying $R\mathbf{p}^* = 0$ and $\sum_i p_i^* = 1$.
Note that $\mathbf{p}^*$ is the (right) eigenvector associated with zero eigenvalue satisfying the probability normalization condition; uniqueness is guaranteed because the system is finite and irreducible as formulated~\cite{Anderson1991}.
Unlike computing the matrix exponential $e^{R t}$, determining $\mathbf{p}^*$ analytically is a feasible task for small $N$.
For $N=5$, we obtain
\begin{equation}
\label{peqN5}
    p_1^* = \frac{k^2}{k^2+10k+5},\quad
    p_2^* = p_3^* = \frac{5k}{k^2+10k+5},\quad
    p_4^* = \frac{5}{k^2+10k+5}.
\end{equation}
In Appendix~\ref{appendix_N6}, we provide the transition matrix $R$ and the equilibrium distribution $\mathbf{p}^*$ for a 6-site system, where six aggregated states are used.

%%%%%%%%%%%%%%%%%%%%%%
\subsection{\label{subsec_surfcov}Equilibrium Surface Coverage}

\subsubsection{Analytic Formulas}

Using the solution $\mathbf{p}(t)$, one can compute the surface coverage $\theta_N(t)$ by a weighted sum of the components of $\mathbf{p}(t)$, where each weight is given as the respective surface coverage that the corresponding configurations represent (i.e.\ $2l_i/N)$.
Since $\mathbf{p^*} = \lim_{t\rightarrow\infty} \mathbf{p}(t)$, the equilibrium coverage $\bar{\theta}_N$ can be obtained as 
%the long-time limit of $\theta_N(t)$, i.e.\ 
$\bar{\theta}_N=\lim_{t\rightarrow\infty}\theta_N(t)$.
Alternatively, we can directly compute $\bar{\theta}_N$ by a weighted sum of the components of $\mathbf{p^*}$.
We thus have 
\begin{equation}
\label{thetaNpi}
    \bar{\theta}_N = \sum_i \frac{2l_i}{N} p_i^*.
\end{equation}
For $N=5$, using the equilibrium distribution of $\mathbf{p}^*$ from Eq.~\eqref{peqN5}, we obtain 
\begin{equation}
    \bar{\theta}_5 = \frac{4k+4}{k^2+10k+5},
\end{equation}
since $(l_1,l_2,l_3,l_4)=(0,1,1,2)$.
Using a similar procedure, we obtain analytic expressions of $\bar{\theta}_N$ for $2\le N\le 8$, see Appendix~\ref{appendix_eqsurfcov}.
These expressions satisfy the following general formulas, depending on whether $N$ is even or odd,
\begin{equation}
\label{thetaeq_evenodd}
    \bar{\theta}_{2m}
    = \ddfrac{\sum_{j=0}^{m-1}\binom{m-1}{j}\binom{m}{j} k^j}{\sum_{j=0}^{m}\binom{m}{j}^2 k^j}, \quad
    \bar{\theta}_{2m+1} 
    = \ddfrac{\sum_{j=0}^{m-1}\binom{2m}{2j+1} k^j}{\sum_{j=0}^{m}\binom{2m+1}{2j+1} k^j}.
\end{equation}
In Section~\ref{subsec_kmc}, we numerically confirm these formulas by performing KMC simulations for larger values of $N$.

\subsubsection{Derivation}

Before investigating various behaviors of $\bar{\theta}_{2m}$ and $\bar{\theta}_{2m+1}$ in Eq.~\eqref{thetaeq_evenodd}, we note that these analytic results can be actually derived by observing that detailed balance is satisfied in the continuous-time Markov chain system.
For any pair of configurations $\alpha$ (with $2l$ atoms) and $\beta$ (with $2l+2$ atoms) that are connected by a certain pair of adsorption and desorption events, their equilibrium probabilities $q_\alpha^*$ and $q_\beta^*$ satisfy $\ra q_\alpha^* = \rd q_\beta^*$, and equivalently, $q_\beta^*/q_\alpha^*=1/k$.
This implies that the equilibrium probability of a configuration depends on the number of adsorbed atoms and is proportional to $k^{-l}$ if there are $2l$ atoms in the configuration.
Hence, the equilibrium probability of the aggregated state $i$ can be written as
\begin{equation}
\label{prob_form}
    p_i^* = c(k)\frac{n_i}{k^{l_i}},
\end{equation}
where $n_i$ is the number of configurations in the state $i$ and $c(k)$ is the normalization constant for $\sum_i p_i^* = 1$.
Using Eqs.~\eqref{thetaNpi} and \eqref{prob_form}, we obtain
\begin{equation}
\label{rewritesum1}
    \bar{\theta}_N = \frac{2 c(k)}{N} \sum_i \frac{l_i n_i}{k^{l_i}}
    = \frac{2 c(k)}{N} \sum_{l=0}^m  \frac{l \tilde{n}_l}{k^l}.
\end{equation}
Note that we rewrote the summation by using index $l$ for possible values of $l_i$ ($l=0,1,\cdots,m$ for both $N=2m$ and $N=2m+1$) and $\tilde{n}_l = \sum_i n_i \delta_{l,l_i}$ (i.e.\ total number of configurations with $2l$ occupied sites) with the Kronecker delta $\delta_{l,l'}$. 
Similarly, we rewrite the normalization condition as
\begin{equation}
\label{rewritesum2}
    1= \sum_i p_i^* = c(k) \sum_i \frac{n_i}{k^{l_i}} = c(k) \sum_{l=0}^m \frac{\tilde{n}_l}{k^l}.
\end{equation}
Therefore, by combining Eqs.~\eqref{rewritesum1} and \eqref{rewritesum2}, we obtain
\begin{equation}
\label{thetaeqgen}
    \bar{\theta}_N = \frac{2}{N} \bigg(\sum_{l=0}^m  \frac{l \tilde{n}_l}{k^l}\bigg) \bigg/ \bigg(\sum_{l=0}^m \frac{\tilde{n}_l}{k^l}\bigg).
\end{equation}
By substituting the expressions of $\tilde{n}_l$ for the odd and even cases, Eqs.~\eqref{ntotnlodd} and \eqref{ntotnleven}, into Eq.~\eqref{thetaeqgen}, one can retrieve $\bar{\theta}_{2m+1}$ and $\bar{\theta}_{2m}$ in Eq.~\eqref{thetaeq_evenodd}.

\subsubsection{Finite System-Size Effect}

\begin{figure}
    \includegraphics[width=\linewidth]{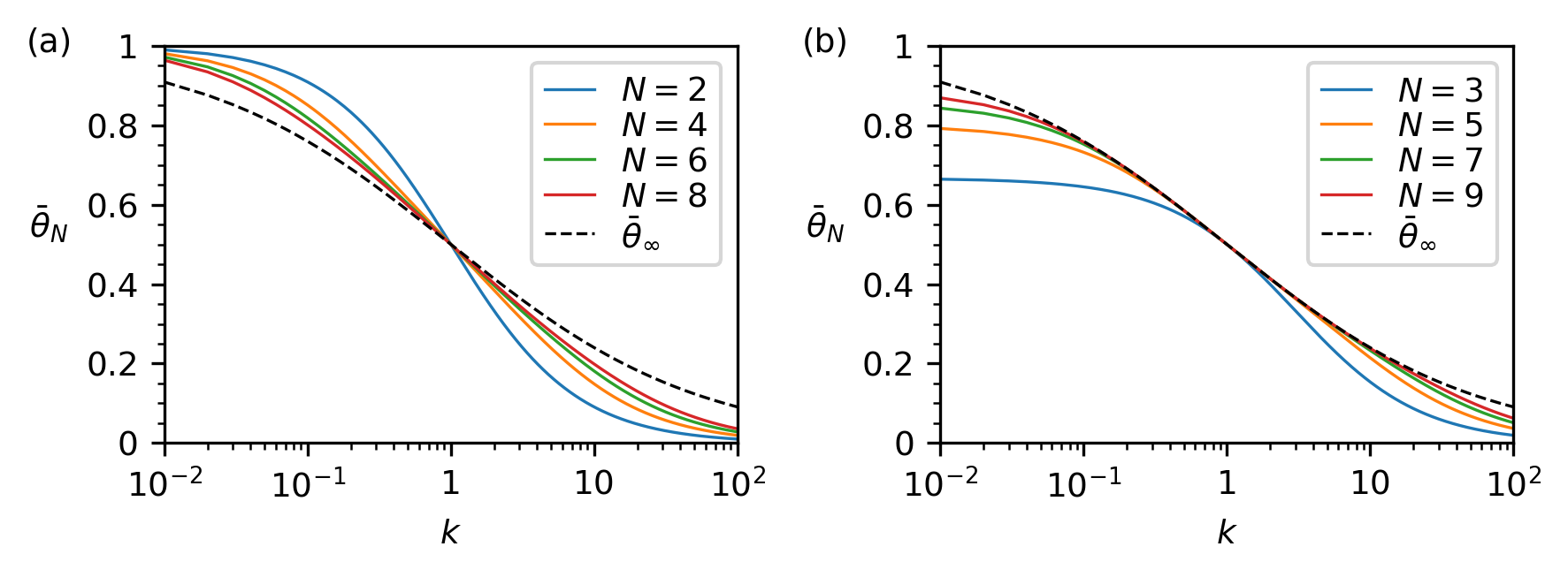}
    \caption{\label{fig_k_eq}
    The equilibrium surface coverage $\bar{\theta}_N$ of an $N$-site system is shown as a function of $k=\rd/\ra$ for various values of $N$.
    The curves of $\bar{\theta}_N$ are plotted for even values of $N$ in panel~(a) and for odd values of $N$ in panel~(b) using Eq.~\eqref{thetaeq_evenodd}.
    To clearly show the finite system-size effect in $\bar{\theta}_N$, the curves with small values of $N$ (solid lines) are compared with the infinite-limit case $\bar{\theta}_\infty$ given in Eq.~\eqref{thetaeq_inf_sec2} (dashed lines).}
\end{figure}

We now analyze the behavior of $\bar{\theta}_{2m}$ and $\bar{\theta}_{2m+1}$.
Figure~\ref{fig_k_eq} shows the curves of $\bar{\theta}_N$ for several small values of $N$ as a function of $k$.
We first confirm that both the even and odd formulas give the same infinite-system limit, i.e.\
$\lim_{m\rightarrow\infty} \bar{\theta}_{2m} = \lim_{m\rightarrow\infty} \bar{\theta}_{2m+1} = \bar{\theta}_\infty$, which coincides with the Langmuir isotherm given in Eq.~\eqref{thetaeq_inf_sec2}.
This shows that, even without infinitely fast surface diffusion of $\Xrm$, each site in the infinite system becomes uncorrelated at equilibrium.
In a finite system, however, reactive sites are not completely uncorrelated, causing finite system-size effects on the equilibrium surface coverage $\bar{\theta}_N$.
We investigate this correlation in Section~\ref{subsec_corr}.

While both the even and odd cases converge to the same value of $\bar{\theta}_\infty$, they exhibit remarkably different convergence behaviors.
Figure~\ref{fig_k_eq} shows that finite system-size effects are more significant and persist longer when $N$ is even.
In the odd case, we observe that for each value of $N$ there is a range of $k$ centered around $k=1$ where the $\bar{\theta}_N$ values are close to $\bar{\theta}_\infty$.
The width of this region increases as $N$ increases.
In the even case, the discrepancy between $\bar{\theta}_N$ and $\bar{\theta}_\infty$ is significantly larger and convergence as $N\rightarrow \infty$ is slower over the entire range of $k$.

\begin{figure}
    \includegraphics[width=\linewidth]{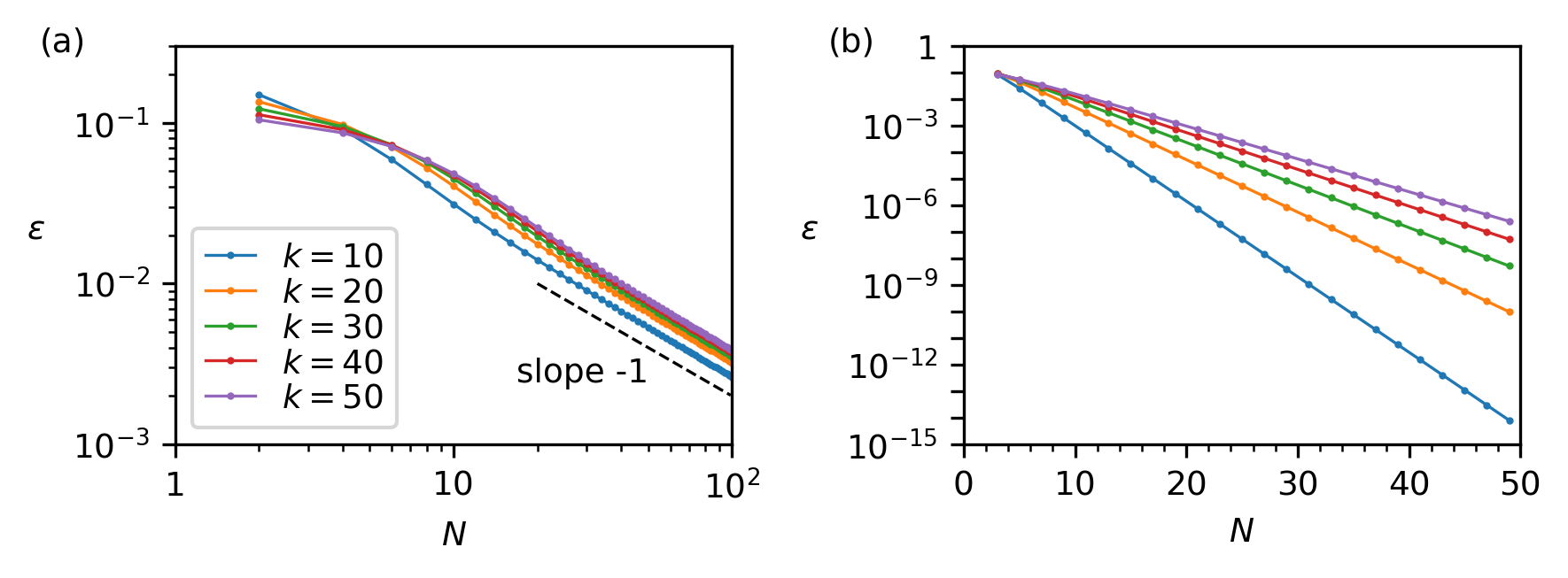}
    \caption{\label{fig_error}
    Convergence behavior of the equilibrium surface coverage $\bar{\theta}_N$ in the limit $N\rightarrow\infty$.
    For even values (shown in panel~(a)) and odd values (panel~(b)) of $N$, the finite system-size effect $\varepsilon(N) = \lvert\bar{\theta}_{N} - \bar{\theta}_{\infty}\rvert$ is plotted.
    Panel~(a): to show $\varepsilon\sim N^{-1}$ for the even case, the curves of $\varepsilon$ are plotted in the $\log$--$\log$ scale for various values of $k$ and a straight line with slope $-1$ is also plotted for comparison. 
    Panel~(b): to show $\varepsilon\sim e^{-a(k)N}$ for the odd case, the curves of $\varepsilon$ are plotted in the semi-log scale for the same values of $k$ as in panel~(a).}
\end{figure}

Figure~\ref{fig_error} shows a more detailed analysis of the finite system-size effect as measured by $\varepsilon = \lvert\bar{\theta}_N - \bar{\theta}_\infty \rvert$. 
As $N$ increases, $\varepsilon$ decreases like $\varepsilon\sim N^{-1}$ in the even case, whereas $\varepsilon$ decreases exponentially (i.e.\ $\varepsilon\sim e^{-a(k)N}$) in the odd case.
Interestingly, we find that the odd case, $\bar{\theta}_{2m+1}$ in Eq.~\eqref{thetaeq_evenodd} is equal to the Pad\'e approximation~\cite{BakerGravesMorris1996} of Eq.~\eqref{thetaeq_inf_sec2} around $k=1$ of orders $(m-1,m)$, meaning that it is the best rational approximation around $k=1$ up to a given order in the power series expansion.
We also observe that the sign and magnitude of the finite system-size effect depend on the value of $k$.
Figure~\ref{fig_k_eq} shows that if $N$ is even then $\bar{\theta}_N > \bar{\theta}_\infty$ (or $\bar{\theta}_N < \bar{\theta}_\infty$) for $k < 1$ (or $k > 1$), while if $N$ is odd then $\bar{\theta}_N < \bar{\theta}_\infty$ for all $k \neq 1$.
When the magnitudes of $\ra$ and $\rd$ are comparable (i.e.\ $k$ is close to unity), finite system-size effects become less significant.
In fact, when $k$ is exactly equal to unity, both formulas match the infinite-limit value $\bar{\theta}_\infty = \frac12$ and thus there is no finite system size effect on $\bar{\theta}_N$.

\subsubsection{Symmetry and Limiting Behaviors}

Figure~\ref{fig_k_eq} also shows that the curve of $\bar{\theta}_{2m}$ has a reflection symmetry around $(k=1,\bar{\theta}=\frac12)$.
In other words, $1-\bar{\theta}_{2m}(\frac1k)=\bar{\theta}_{2m}(k)$ is satisfied, implying that the identical equilibrium surface coverage is obtained when switching the notions of $\Xrm$ (occupied) and $\varnothing$ (unoccupied) and the values of $\ra$ and $\rd$.
However, this symmetry property does not hold for $\bar{\theta}_{2m+1}$.
In fact, if the fully occupied configuration is chosen as the initial configuration, the resulting equilibrium surface coverage $\bar{\theta}_{2m+1}^*$ is different from $\bar{\theta}_{2m+1}$ in Eq.~\eqref{thetaeq_evenodd} and given as $\bar{\theta}_{2m+1}^*(k) = 1-\bar{\theta}_{2m+1}(\frac{1}{k})$.

We also notice that $\bar{\theta}_{2m}$ exhibits correct limiting behaviors for both $k\rightarrow0$ and $k\rightarrow\infty$,
\begin{equation}
\label{thetalimeven}
  \lim_{k\rightarrow 0} \bar{\theta}_{2m} = 1,\quad
  \lim_{k\rightarrow \infty} \bar{\theta}_{2m} = 0,
\end{equation}
whereas $\bar{\theta}_{2m+1}$ shows the correct limiting behavior only for $k\rightarrow\infty$; however, there is a significant system-size effect for $k\rightarrow 0$,
\begin{equation}
\label{thetalimodd}
  \lim_{k\rightarrow 0} \bar{\theta}_{2m+1} = \frac{2m}{2m+1},\quad
  \lim_{k\rightarrow \infty} \bar{\theta}_{2m+1} = 0.
\end{equation}
This system-size effect reflects the fact that for odd $N$ there is always at least one empty site.
Likewise, if the fully occupied configuration is chosen as the initial configuration, a correct limiting behavior is expected for $k\rightarrow 0$ but a significant system-size effect is expected for $k\rightarrow\infty$ because there is always at least one occupied site:
\begin{equation}
\label{thetalimodd2}
  \lim_{k\rightarrow 0} \bar{\theta}_{2m+1}^* = 1,\quad
  \lim_{k\rightarrow \infty} \bar{\theta}_{2m+1}^* = \frac{1}{2m+1}.
\end{equation}
Disparate behaviors shown in the odd and even cases can be related to different accessibility of configurations.
As mentioned in Section~\ref{subsec_conf}, in the odd case, the set of all configurations accessible from the unoccupied configuration is disjoint from the set of all configurations accessible from the fully occupied configuration.
In the even case, the unoccupied and full occupied configurations belong to the same set characterized by $N_+ - N_- = 0$.

%%%%%%%%%%%%%%%%%%%%%%%%%%%%%%%%%%%%%%%%%%%%%%%%%%%%%%%%%%%%%%%
\subsection{\label{subsec_corr}Correlations between Neighboring Sites}

Using the equilibrium probability distribution of accessible configurations, we quantitatively investigate correlations between neighboring sites.
To this end, we first define a random variable $Z_n$ for the occupancy of the $n$th site ($n=1,\cdots,N$), that is, $Z_n=1$ if the $n$th site is occupied and 0 otherwise.
We then define the correlation coefficient $\bar{\rho}_N$ between $Z_1$ and $Z_2$.
Note that any pair of two neighboring sites gives the same result of $\bar{\rho}_N$.
As shown in Appendix~\ref{appendix_rhobarN}, we express $\bar{\rho}_N$ in terms of $[\Xrm]$ and $[\Xrm\Xrm]$:
\begin{equation}
\label{rhoN}
    \bar{\rho}_N 
    = \frac{\mathrm{Cov}[Z_1,Z_2]}{\sqrt{\mathrm{Var}[Z_1]}\sqrt{\mathrm{Var}[Z_2]}}
    = \frac{[\Xrm\Xrm]-[\Xrm]^2}{[\Xrm]\left(1-[\Xrm]\right)},
\end{equation}
where $[\Xrm]=\bar{\theta}_N$ is given in Eq.~\eqref{thetaeq_evenodd} and
\begin{equation}
\label{XXevenodd}
    [\Xrm\Xrm] = \ddfrac{\sum_{j=0}^{m-1}\binom{m-1}{j}^2 k^j}{\sum_{j=0}^m\binom{m}{j}^2 k^j}
    \quad \mbox{for $N=2m$},
    \quad
    [\Xrm\Xrm] = \ddfrac{\sum_{j=0}^{m-1}\binom{2m-1}{2j+1}k^j}{\sum_{j=0}^m\binom{2m+1}{2j+1}k^j}
    \quad \mbox{for $N=2m+1$}.
\end{equation}

\begin{figure}
    \includegraphics[width=\linewidth]{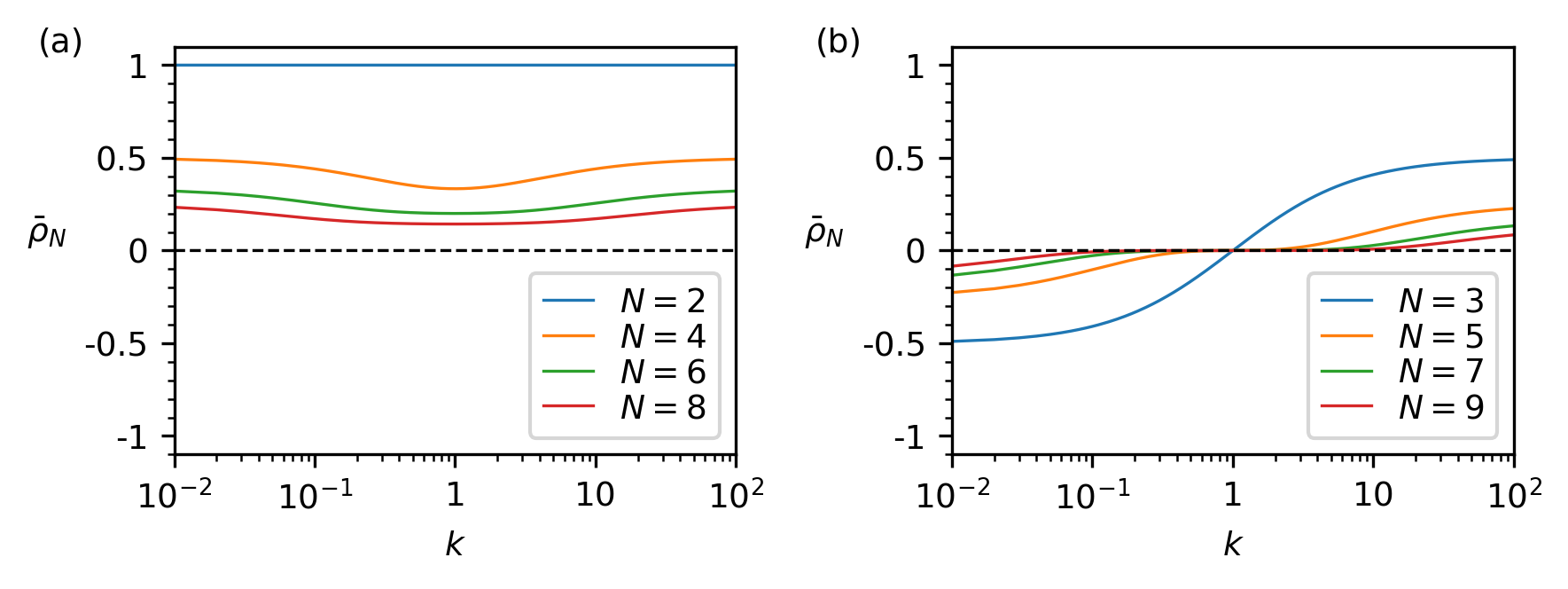}
    \caption{\label{fig_corr}
    The correlation coefficient $\bar{\rho}_N$ between two neighboring sites in an $N$-site system is shown as a function of $k=\rd/\ra$ for various values of $N$.
    The curves of $\bar{\rho}_N$ are plotted by colored solid lines for even values of $N$ in panel~(a) and for odd values of $N$ in panel~(b) using Eq.~\eqref{rhoN}.
    The infinite-limit case (i.e.\ $\bar{\rho}_N\rightarrow 0$ as $N\rightarrow \infty$) is shown by black dashed lines.}
\end{figure}

Figure~\ref{fig_corr} shows the curves of $\bar{\rho}_N$ as a function of $k$ for small values of $N$.
When $N$ is an odd number, $\bar{\rho}_N$ is an odd function in $\log k$ about $k=1$.
Except for $N = 3$, there is a range of $k$ values centered around $k=1$ where the values of $\bar{\rho}_N$ are very close to zero, that becomes wider as $N$ increases.
As $k\rightarrow\infty$, $\bar{\rho}_N \rightarrow -\frac{1}{N-1}$ as $k\rightarrow0$ and $\bar{\rho}_N \rightarrow \frac{1}{N-1}$. 
For an even number $N>2$, as in the odd case, there is a range of $k$ values around $k=1$ where the magnitude of $\bar{\rho}_N$ becomes smaller that increases as $N$ increases.
However, for the even case, $\bar{\rho}_N$ is an even function in $\ln k$ about $k=1$, is positive for all values of $k$, and has the minimum value $\frac{1}{N-1}$ at $k=1$.
As $k\rightarrow0$ or $k\rightarrow\infty$, $\bar{\rho}_N \rightarrow \frac{2}{N}$.

The behavior of the correlation coefficient $\bar{\rho}_N$ for the even and odd cases explain why the finite system-size effect on the equilibrium surface coverage becomes more significant and persistent in the even case.
In the odd case, correlations between neighboring sites are negligible in equilibrium in a neighborhood of
$k=1$, in contrast to the even case where they are not.
Furthermore, the range of $k$ with negligible correlation increases as $N$ increases.
We revisit this relation of the finite system-size effect and correlations between neighboring sites in Section~\ref{subsec_size_surfdiff}, where we consider surface diffusion.

%%%%%%%%%%%%%%%%%%%%%%%%%%%%%%%%%%%%%%%%%%%%%%%%%%%%%%%%%%%%%%%
\subsection{\label{subsec_kmc}Time-Transient Behavior of $\theta_N(t)$}

By performing KMC simulations, we numerically validate our analytic results for the equilibrium surface coverage $\bar{\theta}_N$ given in Eq.~\eqref{thetaeq_evenodd}, and also observe the time-transient behavior of $\theta_N(t)$.
For the setup of KMC simulations, see the Supplementary Material.

\begin{figure}
    \includegraphics[width=\linewidth]{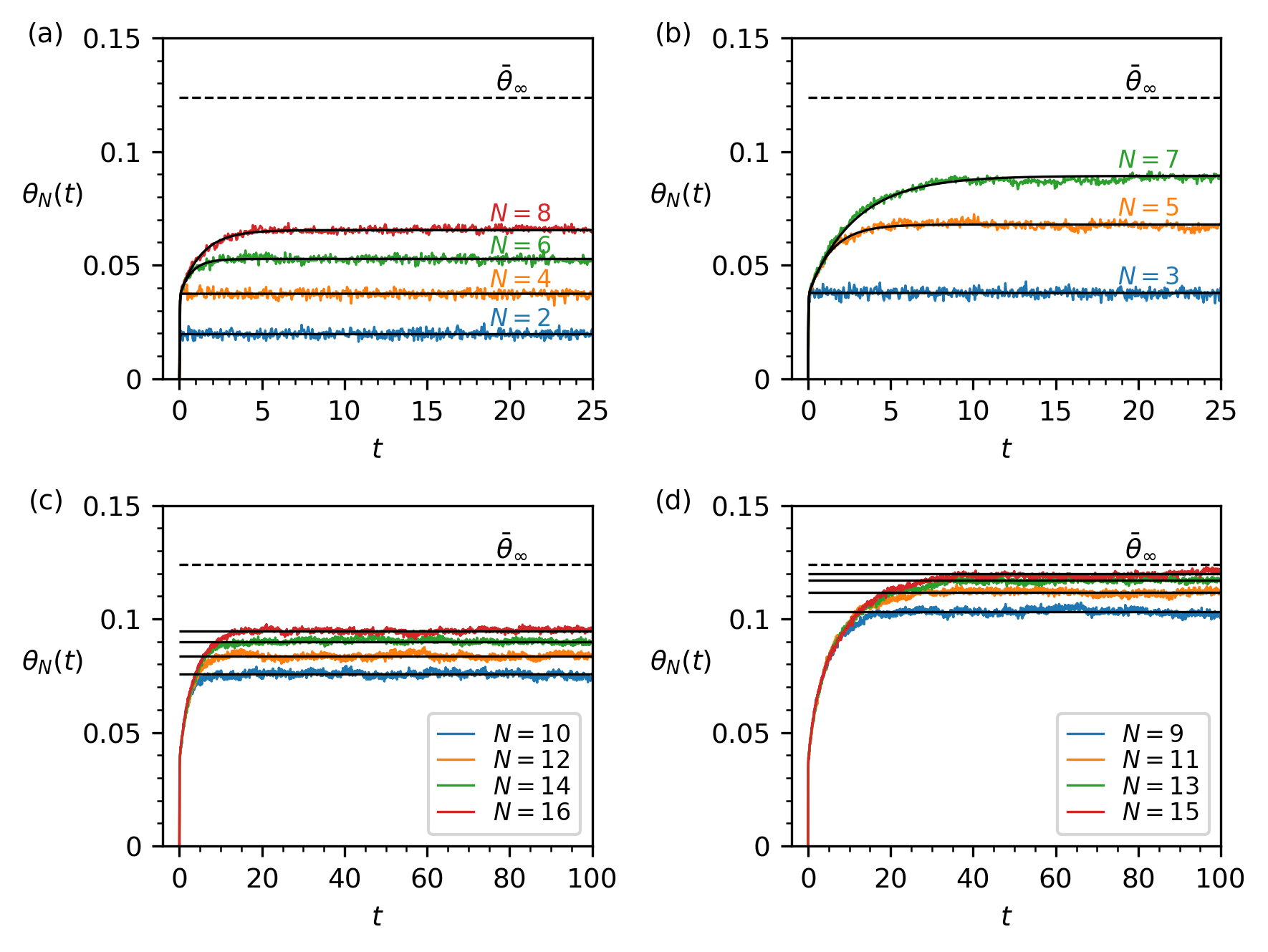}
    \caption{\label{fig_kmc_val}For a large value of $k=50$ ($\ra=1$ and $\rd=50$), time-dependent surface coverages $\theta_N(t)$ obtained by KMC simulations (colored lines) are compared with the CME results (black solid lines): for $2\le N\le 8$, the even case in panel~(a) and the odd case in panel~(b); 
    for $9 \le N \le 16$, the even case in panel~(c) and the odd case in panel~(d).
    %For the CME results, 
    The curves of $\theta_N(t)$ computed by the CME are plotted in panels~(a) and (b). For larger systems shown
    in panels~(c) and (d), where CME results are not available, we plot the values of $\bar{\theta}_N$ computed by Eq.~\eqref{thetaeq_evenodd}.
    The infinite-system equilibrium coverage $\bar{\theta}_\infty$ is also shown by the dashed line.
    Error bars for KMC simulations are not shown for visual clarity.
    The magnitude of error bars is comparable to that of fluctuations appearing in each curve.}
\end{figure}

In Figure~\ref{fig_kmc_val}, we show the KMC and CME results for a large value of $k=50$.
Panels~(a) and (b) show the time profiles of $\theta_N(t)$ for small values of $N$ up to 8.
Since the transition matrix $R$ can be explicitly given for these $N$ values, $\theta_N (t)$ can be also obtained by numerically solving the CME~\eqref{cme_mform}.
We first confirm the agreement between the KMC and CME results, which cross-validates both approaches.
Due to the large value of $k$, we observe significant system-size effects on the long-time limit of $\theta_N(t)$ (i.e.\ $\bar{\theta}_N$).
In addition, these effects are more severe when $N$ is even.
For example, the result with $N=8$ has larger system-size effects than $N=5$. 
We also investigate how these effects develop as time increases.
Early in the simulation $\theta_N(t)$ grows rapidly and its curves with different values of $N$ coincide. 
As later times, however, curves with smaller values of $N$ start to reach their equilibrium values and diverge from curves corresponding to larger values of $N$, reflecting  a lower long-time limit value $\bar{\theta}_N = \lim_{t\rightarrow\infty} \theta_N(t)$ for smaller $N$.
This behavior appears in both even and odd cases.
Panels (c) and (d) show the time profiles of $\theta_N (t)$ for larger values of $N$, $9\le N\le 16$ as obtained by KMC.
As expected, those curves converge to the values $\bar{\theta}_N$ predicted by Eq.~\eqref{thetaeq_evenodd}, which confirms the validity of these analytic results. 
The characteristic behaviors appearing in panels~(a) and (b) are also observed.
In particular, remarkably slow convergence of $\theta_N (t)$ to $\theta_\infty (t)$ is observed for even values of $N$.
In addition, for smaller values of $N$, $\bar{\theta}_N$ is smaller as a result of the earlier rollover of the $\theta_N(t)$ curves compared to larger values of $N$ for both even and odd cases.

%%%%%%%%%%%%%%%%%%%%%%%%%%%%%%%%%%%%%%%%%%%%%%%%%%%%%%%%%%%%%%%
\section{\label{sec_surfdiff}Effect of Surface Diffusion}

We now consider the case where the lattice system undergoes not only reversible dissociative adsorption but also surface diffusion.
In other words, we allow an absorbed $\Xrm$ to hop into a neighboring site if the site is unoccupied as described by
\begin{equation}
\label{surfdiff}
    \Xrm\varnothing \xrightleftharpoons[\rdiff]{\rdiff} \varnothing\Xrm,
\end{equation}
where the rate is denoted by $\rdiff$.

Including surface diffusion is expected to reduce correlations between neighboring sites that are caused by reversible dissociative adsorption.
Hence, while each surface diffusion event itself does not change the instantaneous value of surface coverage, the surface coverage dynamics is modified by surface diffusion. 
In this section, we investigate how surface diffusion affects the finite system-size effect on surface coverage dynamics.

\begin{figure}
    \includegraphics[width=0.5\linewidth]{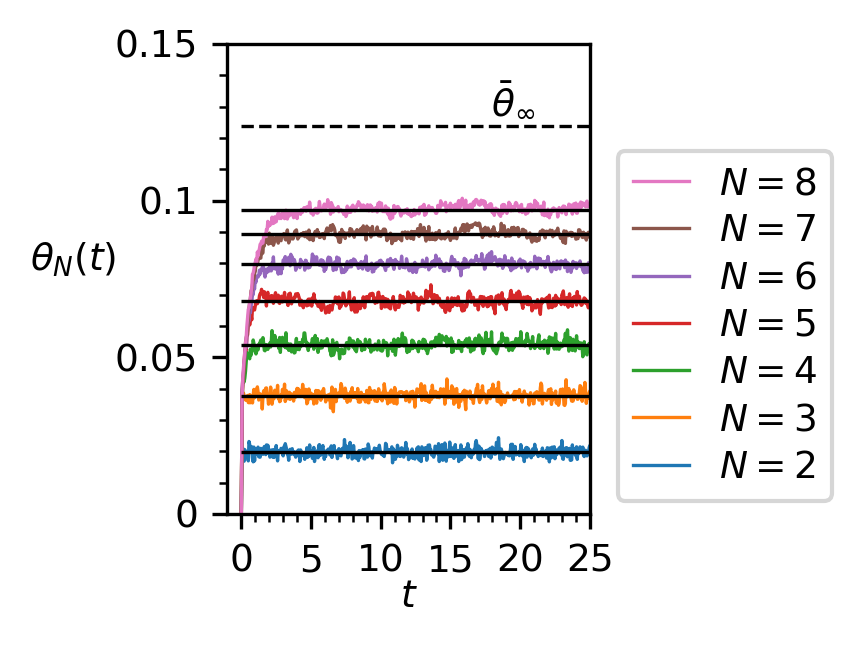}
    \caption{\label{fig_kmc_surfdiff}
    For a lattice system undergoing reversible dissociative adsorption as well as surface diffusion, time-dependent surface coverages $\theta_N(t)$ obtained by KMC simulations (colored lines) are shown for small $N$ values, $2\le N\le 8$.
    The value of $k$ is set to 50 using $\ra=1$ and $\rd=50$, whereas the rate for surface diffusion is set to $\rdiff=1$.
    The black solid lines indicate the positions of the equilibrium surface coverage $\bar{\theta}_N$ estimated by the analytic formulas in Eq.~\eqref{theta_eq_surfdiff} for each $N$.
    The infinite-system equilibrium coverage $\bar{\theta}_\infty$ is also shown by the dashed line.
    Note that the simulation results without surface diffusion are shown in Figure~\ref{fig_kmc_val}.}
\end{figure}

Before discussing analytic results, we first present KMC simulation results to emphasize different behaviors of $\bar{\theta}_N$ and $\theta_N (t)$ when surface diffusion is considered. 
Figure~\ref{fig_kmc_surfdiff} shows the time profiles of $\theta_N(t)$ for small values of $N$ up to 8.
The value of $k$ is set to 50 using $\ra=1$ and $\rd=50$ whereas the rate for surface diffusion is set to $\rdiff=1$.
In contrast with the no-diffusion case shown in Figure~\ref{fig_kmc_val} the positions of the equilibrium surface coverage $\bar{\theta}_N$ are in order (i.e.\ $\bar{\theta}_2<\bar{\theta}_3<\cdots<\bar{\theta}_8<\bar{\theta}_\infty$).
In other words, the significant finite system-size effect in the even case that appears in the no-diffusion case is absent.
We also observe that, for each value of $N$, $\theta_N(t)$ reaches its equilibrium value $\bar{\theta}_N$ faster due to surface diffusion.

\begin{figure}
    \includegraphics[width=\linewidth]{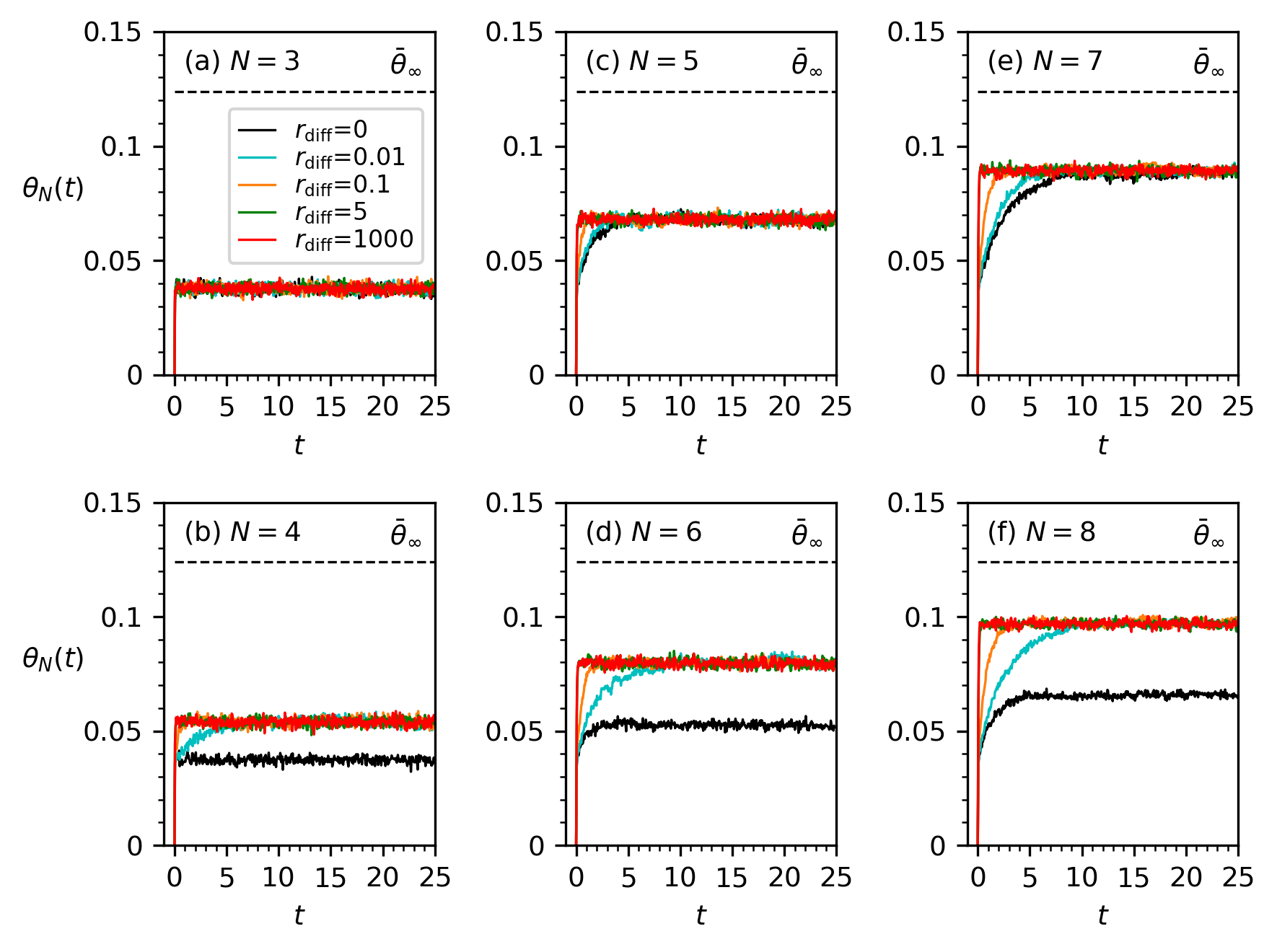}
    \caption{\label{fig_Finite_diff_N}
    For each system size $N$ ($3\le N\le 8$), the time profiles of the surface coverage $\theta_N (t)$ obtained from KMC simulations with $k=50$ ($\ra=1$, $\rd=50$) and different values of the surface-diffusion rate $\rdiff$ are plotted.
    The infinite-system equilibrium coverage $\bar{\theta}_\infty$ is also shown by the dashed line.
    Note that surface diffusion cannot be considered for $N=2$ and thus this case is omitted.}
\end{figure}

In order to further investigate these behaviors, for each value of $N$ ($3\le N\le 8$), we compute the time profiles of $\theta_N(t)$ for different values of $\rdiff$ and compare them with the no-diffusion case (i.e.\ $\rdiff=0$) in Figure~\ref{fig_Finite_diff_N}.
For each odd value of $N$, the equilibrium $\bar{\theta}_N$ is the same for all values of $\rdiff$, including zero.
The main difference due to the $\rdiff$ value is that $\theta_N(t)$ reaches $\bar{\theta}_N$ faster as $\rdiff$ increases.
For the even case, the same observations are made for all nonzero values of $\rdiff$.
However, the no-diffusion case with $\rdiff=0$ is singular in the sense that its equilibrium value is different from that obtained from all nonzero values of $\rdiff$.

The singular behavior of the no-diffusion case for an even value of $N$ results from an insufficient number of accessible configurations, as discussed earlier.
For an odd value of $N$, the total number of configurations accessible from the initially unoccupied state is equal to $n_\mathrm{tot} = 2^{N-1}$ whether surface diffusion is included or not.
On the contrary, for an even value of $N$, if surface diffusion is not included, some configurations become inaccessible due to the conserved quantity $N_+-N_-=0$ and $n_\mathrm{tot}=\binom{N}{N/2}$, see Eq.~\eqref{ntotnleven}.
If surface diffusion is included, however, all configurations with an even number of occupied sites become accessible.
Hence, for both $N=2m$ and $N=2m+1$, we have
\begin{equation}
\label{ntotnl_surfdiff}
    \tilde{n}_l = \binom{N}{2l}, \quad 
    n_\mathrm{tot} = \sum_{l=0}^{m} \tilde{n}_l = 2^{N-1}.
\end{equation}

Based on this observation, we derive analytic expressions for $\bar{\theta}_N$ for the surface-diffusion case.
Under the assumption that detailed balance also holds in the presence of surface diffusion, Eq.~\eqref{thetaeqgen} is valid.
By substituting Eq.~\eqref{ntotnl_surfdiff} to Eq.~\eqref{thetaeqgen}, we obtain
\begin{equation}
\label{theta_eq_surfdiff}
    \bar{\theta}_{2m}
    = \ddfrac{\sum_{j=0}^{m-1}\binom{2m-1}{2j} k^j}{\sum_{j=0}^{m}\binom{2m}{2j} k^j},\quad
    \bar{\theta}_{2m+1} 
    = \ddfrac{\sum_{j=0}^{m-1}\binom{2m}{2j+1} k^j}{\sum_{j=0}^{m}\binom{2m+1}{2j+1} k^j}.
\end{equation}
Similarly, we obtain analytic expressions of the correlation coefficient $\bar{\rho}_N$ using Eqs.~\eqref{rhoN}, \eqref{theta_eq_surfdiff} and
\begin{equation}
\label{XXevenodd_surfdiff}
    [\Xrm\Xrm] = \ddfrac{\sum_{j=0}^{m-1}\binom{2m-2}{2j}k^j}{\sum_{j=0}^m\binom{2m}{2j}k^j}
    \quad \mbox{for $N=2m$},
    \quad
    [\Xrm\Xrm] = \ddfrac{\sum_{j=0}^{m-1}\binom{2m-1}{2j+1}k^j}{\sum_{j=0}^m\binom{2m+1}{2j+1}k^j}
    \quad \mbox{for $N=2m+1$}.
\end{equation}
See Appendix~\ref{appendix_rhobarN} for the derivation of Eq.~\eqref{XXevenodd_surfdiff}.
We note that, in the odd case, the analytic expressions of $\bar{\theta}_N$ and $\bar{\rho}_N$ are exactly the same as the no-diffusion case, see Eqs.~\eqref{thetaeq_evenodd} and \eqref{XXevenodd}, and this is why the singular behavior does not appear in the odd case in Figure~\ref{fig_Finite_diff_N}.

\begin{figure}
    \includegraphics[width=\linewidth]{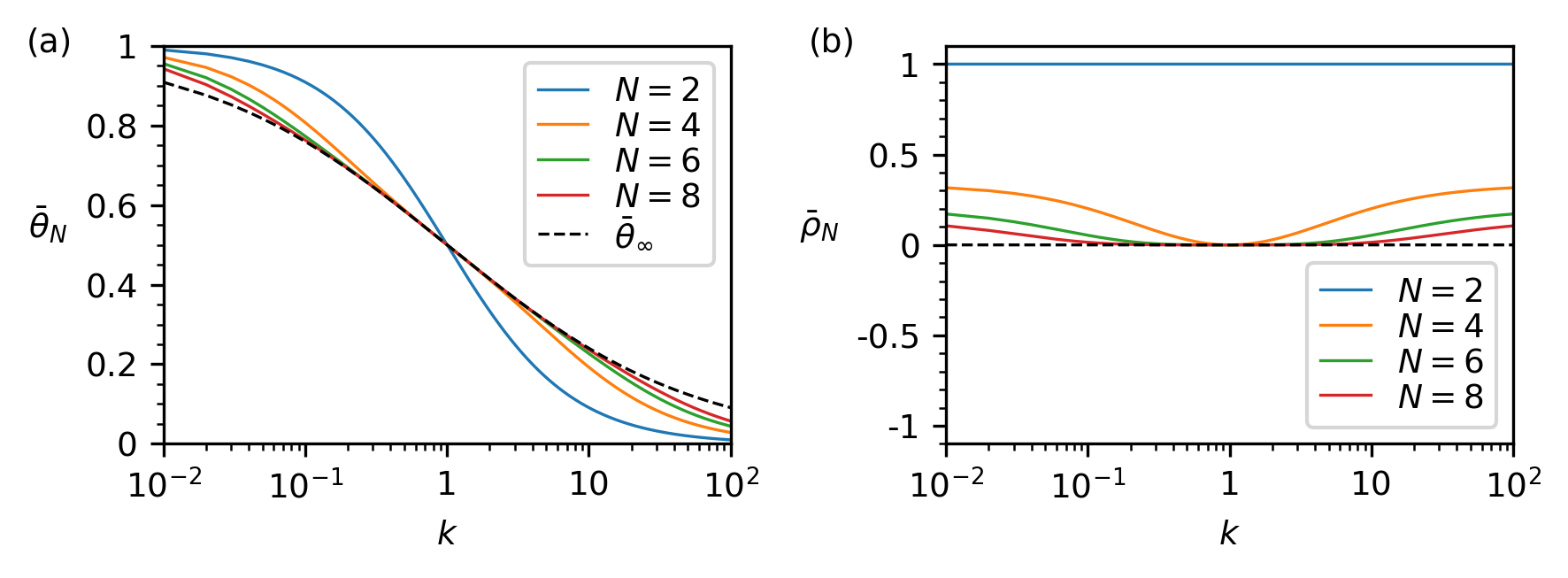}
    \caption{\label{fig_Deven}In panel~(a), the equilibrium surface coverage $\bar{\theta}_N$ with surface diffusion is shown as a function of $k$ for small even numbers $N$ using Eq.~\eqref{theta_eq_surfdiff}.
    The infinite-system limit $\bar{\theta}_\infty$ is also shown.
    In panel~(b), the correlation coefficient $\bar{\rho}_N$ with surface diffusion is shown for small even numbers $N$ using Eqs.~\eqref{rhoN}, \eqref{theta_eq_surfdiff}, and \eqref{XXevenodd_surfdiff}.
    The infinite-system limit (i.e.\ $\bar{\rho}_\infty\rightarrow 0$ as $N\rightarrow\infty$) is also shown.
    The odd case is omitted because it is exactly same as the no-diffusion case, see Figure~\ref{fig_k_eq}(b) and Figure~\ref{fig_corr}(b).}
\end{figure}

Figure~\ref{fig_Deven} shows the curves of $\bar{\theta}_N$ and $\bar{\rho}_N$ versus $k$ for small even values $N$ when surface diffusion is considered.
We note that the odd case is exactly the same as the no-diffusion case shown in Figures~\ref{fig_k_eq} and \ref{fig_corr}.
Contrary to the no-diffusion case, for each even value of $N$, there is a range of $k$ values where $\bar{\theta}_N$ is much closer to $\bar{\theta}_\infty$.
This explains why significant system-size effects observed in Figure~\ref{fig_corr} for the even case do not appear here.
In fact, for both even and odd cases, it is observed that $\varepsilon=\left|\bar{\theta}_N-\bar{\theta}_\infty\right|$ decreases exponentially, i.e.\ $\varepsilon\sim e^{-a(k)N}$;
the convergence plot is similar to Figure~\ref{fig_error}(b) (see the Supplementary Material).
We also notice that, in the plot of the correlation coefficient $\bar{\rho}_N$ in Figure~\ref{fig_Deven}(b), for each even value of $N>2$ there is a corresponding range of $k$ where $\bar{\rho}_N$ is much closer to zero.
This demonstrates the close relation between the finite system-size effect on the equilibrium surface coverage and correlations between neighboring sites.

\section{\label{sec_conclusion}Conclusion}

We have considered the surface coverage dynamics where reversible dissociative adsorption occurs on an initially unoccupied linear lattice.
Unlike the molecular (or non-dissociative) adsorption case, this system exhibits finite system-size effects caused by dynamic correlations between neighboring sites.
We investigated this finite size effect on the equilibrium surface coverage and relate it to non-vanishing static site correlations introduced by reversible dissociative adsorption.
We also investigated the effects of surface diffusion of adsorbed atoms, which reduces site correlations. 

We modeled the equilibrium surface coverage $\bar{\theta}_N$ and time-transient surface coverage $\theta_N(t)$ of a finite lattice with $N$ reactive sites using the chemical master equation (CME) and kinetic Monte Carlo (KMC).
We derived analytical expressions for $\bar{\theta}_N$  and verified them numerically for the case without surface diffusion, see Eq.~\eqref{thetaeq_evenodd}, and the case with surface diffusion, see Eq.~\eqref{theta_eq_surfdiff}.
Without surface diffusion, finite system-size effects are significant for even $N$ when the ratio $k=\rd/\ra$ is much larger or smaller than unity.
By comparing to the case with surface diffusion, the behavior observed for the even case $\bar{\theta}_{2m}$ in Eq.~\eqref{thetaeq_evenodd} was explained by an insufficient number of accessible configurations.
We further related this behavior with the persistent positive correlation coefficient $\bar{\rho}_N$ for even $N$, and demonstrated the close relation between the finite system-size effect on the equilibrium surface coverage and correlations between neighboring sites.

We draw the reader's attention to the following points.
First, in our study, the equilibrium was defined as the steady state that the system attains with a given initial configuration as opposed to the one defined via a grand canonical distribution of configurations.
Our analysis relies on the fact that not every pair of configurations is mutually accessible via reversible dissociative adsorption and, as a result, configurations are partitioned into classes such that only configurations in the same class are accessible to each other.
In the sense that the steady-state of the system depends on the initial state, the system is not ergodic. 
Second, as discussed in Ref.~\citenum{LiuEvans2015}, the conservation of the quantity $N_+ - N_-$ still holds in a 2D square lattice.
Hence, a similar system-size effect is expected when a square lattice has an even number of sites in each direction.
Third, in a real system, the range of $k=\rd/\ra$ can be much wider than the range $[10^{-2},10^2]$ considered in our study.
While the considered range roughly corresponds to a range of $[-0.1,0.1]$ eV for adsorption free energies at room temperature, a model with strong binding can easily have a value beyond this range~\cite{GelssMateraSchuette2016}.
Hence, more significant finite-size effects can be expected.
Although this system-size effect may not be significant when surface diffusion or other surface reactions are introduced, our study implies that caution should be exercised when lattice KMC modeling is used for a surface lattice system undergoing reversible dissociative adsorption of dimers.

This study has the following possible future directions.
First, one can investigate non-periodic systems or two-dimensional lattice systems.
Alternatively, since our model can be considered as a simple theoretical model for active sites adjacent to doped sites~\cite{HakalaKronbergLaasonen2017}, one can also consider a system consisting of several strips where strip $k$ has $N_k$ sites and $N_k$ follows, for example, a Poisson distribution.
Second, one can also consider dissociative adsorption of heteronuclear diatomic molecules, e.g.\ $\mathrm{NO}$~\cite{AbadBoehmeRoman2007}.
Third, as mentioned in Introduction, our findings will be useful for the development of a multiscale simulation method for a fluid-solid interfacial system, where KMC is coupled with a mesoscopic continuum method, for example, fluctuating hydrodynamics~\cite{ZarateSengers2006, SrivastavaLadigesNonakaGarciaBell2023}.

\section*{Supplementary Material}

KMC simulation setup; Convergence behavior of $\bar{\theta}_N$ in the presence of surface diffusion.

\acknowledgments

C.K. thanks Drs.\ Fran\c{c}ois Blanchette and Lei, Yue (both at UC Merced) for helpful discussions on the combinatorial expressions for the number of accessible configurations.
This work was supported in part by the National Science Foundation under Grant No.\ CHE-2213368 and Grant No.\ DMS-1840265. 
This work was also supported in part by the U.S.\ Department of Energy, Office of Science, Office of Advanced Scientific Computing Research, Applied Mathematics Program under Contract No.\ DE-AC02-05CH11231.

\appendix

\section{\label{appendix_N6}Transition Matrix and Equilibrium Distribution for a 6-Site System}

For $N=6$ there are 20 accessible configurations and we group them into the following 6 aggregated states: $\mathbb{C}(\{\varnothing\varnothing\varnothing\varnothing\varnothing\varnothing\})$ (1 configuration), $\mathbb{C}(\{\Xrm\Xrm\varnothing\varnothing\varnothing\varnothing\})$ (6 configurations), 
$\mathbb{C}(\{\Xrm\varnothing\varnothing\Xrm\varnothing\varnothing\})$ (3 configurations), 
$\mathbb{C}(\{\Xrm\Xrm\Xrm\Xrm\varnothing\varnothing\})$ (6 configurations), 
$\mathbb{C}(\{\Xrm\Xrm\varnothing\Xrm\Xrm\varnothing\})$ (3 configurations),
$\mathbb{C}(\{\Xrm\Xrm\Xrm\Xrm\Xrm\Xrm\})$ (1 configuration).
The transition matrix $R$ is given as
\begin{equation}
    R = \begin{bmatrix} -6\ra & \rd & 0 & 0 & 0 & 0\\
    6\ra & -3\ra-\rd & 0 & 2\rd & 2\rd & 0 \\
    0 & 0 & -2\ra & \rd & 0 & 0\\
    0 & 2\ra & 2\ra & -\ra-3\rd & 0 & 6\rd \\
    0 & \ra & 0 & 0 & -2\rd & 0 \\
    0 & 0 & 0 & \ra & 0 & -6\rd
    \end{bmatrix}.
\end{equation}
The equilibrium probability distribution $\mathbf{p}^*$ is given as
\begin{equation}
\begin{split}
    &p_1^* = \frac{k^3}{k^3+9 k^2+9k+1}, \quad
    p_2^* = \frac{6 k^2}{k^3+9 k^2+9k+1}, \quad
    p_3^* = \frac{3 k^2}{k^3+9 k^2+9k+1}, \quad \\
    &p_4^* = \frac{6 k}{k^3+9 k^2+9k+1}, \quad
    p_5^* = \frac{3 k}{k^3+9 k^2+9k+1}, \quad
    p_6^* = \frac{1}{k^3+9 k^2+9k+1}.
\end{split}
\end{equation}

\section{\label{appendix_eqsurfcov}Equilibrium Surface Coverage for Small Systems}

The equilibrium surface coverage of an $N$-site system, $\bar{\theta}_N$, is expressed in terms of $k=\rd/\ra$.
The following formulas are derived analytically from the corresponding $N$-site CME:
\begin{subequations}
\label{specificthetaeq}
\begin{align}
    \bar{\theta}_2 &= \frac{1}{k+1}, \\
    \bar{\theta}_3 &= \frac{2}{k+3}, \\
    \bar{\theta}_4 &= \frac{2k+1}{k^2+4k+1}, \\
    \bar{\theta}_5 &= \frac{4k+4}{k^2+10k+5}, \\
    \bar{\theta}_6 &= \frac{3k^2+6k+1}{k^3+9k^2+9k+1}, \\
    \bar{\theta}_7 &= \frac{6k^2+20k+6}{k^3+21k^2+35k+7}, \\
    \bar{\theta}_8 &= \frac{4k^3+18k^2+12k+1}{k^4+16k^3+36k^2+16k+1}.
\end{align}
\end{subequations}

\section{\label{appendix_rhobarN}Derivation of $\bar{\rho}_N$ and $[\Xrm\Xrm]$}

The correlation coefficient of $Z_1$ and $Z_2$ is defined as
\begin{equation}
\label{barhoNdef}
    \bar{\rho}_N = \frac{\mathrm{Cov}[Z_1,Z_2]}{\sqrt{\mathrm{Var}[Z_1]}\sqrt{\mathrm{Var}[Z_2]}}.
\end{equation}
We express $\bar{\rho}_N$ in terms of $[\Xrm]$ and $[\Xrm\Xrm]$.
By noting $Z_1^2=Z_1$ (whether $Z_1$ has 0 or 1), we first obtain
\begin{equation}
\label{varZ1}
    \mathrm{Var}[Z_1] 
    = \mathbb{E}[Z_1^2] - \left(\mathbb{E}[Z_1]\right)^2
    = \mathbb{E}[Z_1] - \left(\mathbb{E}[Z_1]\right)^2
    = \mathbb{E}[Z_1]\left( 1 - \mathbb{E}[Z_1]\right)
    = [\Xrm]\left(1-[\Xrm]\right),
\end{equation}
and similarly $\mathrm{Var}[Z_2] = [\Xrm]\left(1-[\Xrm]\right)$.
We also obtain
\begin{equation}
    \mathrm{Cov}[Z_1,Z_2] 
    = \mathbb{E}[Z_1 Z_2] - \mathbb{E}[Z_1] \mathbb{E}[Z_2]
    = [\Xrm\Xrm]-[\Xrm]^2,
\end{equation}
and thus obtain Eq.~\eqref{rhoN}.

We derive analytic expressions for $[\Xrm\Xrm]$ given in Eq.~\eqref{XXevenodd}.
Following a similar procedure to obtain Eq.~\eqref{thetaeqgen}, we first obtain 
\begin{equation}
\label{XXgen}
    [\Xrm\Xrm] = c(k) \sum_{l=0}^m \frac{\tilde{n}_{l,\Xrm\Xrm}}{k^l}
    = \bigg(\sum_{l=0}^m \frac{\tilde{n}_{l,\Xrm\Xrm}}{k^l}\bigg) \bigg/
    \bigg(\sum_{l=0}^m \frac{\tilde{n}_l}{k_l}\bigg).
\end{equation}
Here, $\tilde{n}_{l,\Xrm\Xrm}$ denotes the number of configurations with $2l$ occupied sites, the first two sites of which are $\Xrm\Xrm$.
For $N=2m+1$, counting the number of configurations with $2l$ occupied sites that start with $\Xrm\Xrm$ is equivalent to counting the number of ways to choose $2l-2$ items from $2m-1$ items:
\begin{equation}
\label{nlXXodd}
    \tilde{n}_{l,\Xrm\Xrm} = \binom{2m-1}{2l-2}\quad\mbox{for  $l=1,2,\cdots,m$}.
\end{equation}
For $N=2m$, $\tilde{n}_{l,\Xrm\Xrm}$ is equal to the number of accessible configurations with $2(l-1)$ occupied sites for a finite system with $2(m-1)$ sites:
\begin{equation}
\label{nlXXeven}
    \tilde{n}_{l,\Xrm\Xrm} = \binom{m-1}{l-1}^2\quad\mbox{for  $l=1,2,\cdots,m$}. 
\end{equation}
By substituting Eqs.~\eqref{nlXXodd}--\eqref{nlXXeven} and \eqref{ntotnlodd}--\eqref{ntotnleven} into Eq.~\eqref{XXgen}, one can obtain Eq.~\eqref{XXevenodd}.

In the surface-diffusion case, for both $N=2m$ and $N=2m+1$, $\tilde{n}_{l,\Xrm\Xrm}$ is given as
\begin{equation}
\label{nlXX_surfdiff}
    \tilde{n}_{l,\Xrm\Xrm} = \binom{N-2}{2l-2}\quad\mbox{for  $l=1,2,\cdots,m$}.
\end{equation}
By substituting Eqs.~\eqref{ntotnl_surfdiff} and \eqref{nlXX_surfdiff} into Eq.~\eqref{XXgen}, we obtain Eq.~\eqref{XXevenodd_surfdiff}.

%\bibliography{refs}

\bibliography{manuscript.bbl}

\end{document}